\documentstyle[pre,aps,tighten,floats,epsf]{revtex}
\begin{document}
\renewcommand
\baselinestretch{2}
\large
\title{Energetics of rocked inhomogeneous ratchets}
\author{Debasis Dan and A. M. Jayannavar}
\address{Institute of Physics, Sachivalaya Marg,
Bhubaneswar 751005, India}
\maketitle
\begin{abstract}
             We study the efficiency of frictional thermal ratchets 
driven by finite frequency driving force and in contact with a 
heat bath. The efficiency
exhibits varied behavior with driving frequency. Both
nonmonotonic and monotonic behavior have been observed. In particular 
the magnitude of efficiency in finite frequency regime may be more than
the efficiency in the adiabatic regime. This is our
central result for rocked ratchets. We also
show that for the simple potential we have chosen, the presence of
only spatial  
asymmetry (homogeneous system) or only frictional ratchet (symmetric
potential profile), the adiabatic efficiency
is always more than in the nonadiabatic case. 
\end{abstract}
\newpage
   
                   Much has been studied in ratchet models (Brownian
motors) to determine how directed motion appears out of nonequilibrium
fluctuations in the absence of any net macroscopic force. Here 
athermal fluctuation combined with spatial or temporal anisotropy 
conspire to generate systematic motion even in the absence of net bias~\cite{reimann}. These
studies have been inspired by the observations on molecular 
motors in biological systems~\cite{ajdari}. To this effect several
physical  models have been proposed under the
name of rocking ratchets, flashing
ratchets, diffusion ratchets, correlation
ratchets, frictional ratchets ~\cite{reimann} etc. In most of these 
systems, focus was mainly
on the behavior of probability current with change in system
parameter like temperature, amplitude of external force, correlation
time, etc. The
efficiency with which these ratchets convert 
fluctuation to useful work is a subject of much recent
interest ~\cite{seki,taka1}. 
New questions regarding the nature of heat engines (reversible or irreversible)
at molecular scales are being investigated. Especially the source of
irreversibility and whether the irreversibility can be suppressed such that 
efficiency can approach that of Carnot cycle 
~\cite{ajdari,hondu_etall} and generalization of 
thermodynamics principles to nonequilibrium steady state are being
investigated ~\cite{hatano}. 
We use the method of stochastic energetics developed by Sekimoto~\cite{seki}. 
In this scheme, quantities like heat, work done and input energy can be 
calculated within the framework of Langevin equation. 
Using this approach efficiency has been studied mainly as a function of 
temperature and load in rocking, oscillating and frictional 
ratchets. In some cases it has been shown that efficiency can be maximized 
at finite temperature~\cite{taka1,Dan3}.
The efficiency in these systems are rather small, the 
reason being inherent irreversibility of these engines due to 
finite current. 

          In our present work we mainly explore the nature of efficiency in  
frictional rocking ratchets as function of 
frequency of external drive.    
The systematic study of efficiency as a function
of frequency in rocked ratchets has not been studied so far. 
We show in the following that a rocking ratchet with
inhomogeneous friction coefficient can have efficiency which is a
nonmonotonic function of frequency. In some parameter range, \textbf{the 
efficiency in the nonadiabatic regime can even be larger than in the adiabatic 
regime.} This is solely due
to the interplay between the asymmetry in the potential 
and the space dependent friction coefficient. In absence of frictional 
inhomogeneity our system reduces to a conventional rocked ratchet.
It may also happen that
inspite of this nonmonotonic behavior with frequency the adiabatic
efficiency is larger than the nonadiabatic efficiency. This shows that
in the nonequilibrium regime efficiency exhibits
complex behavior some which are against the established
tenets of equilibrium phenomena, like efficiency in
quasi-static processes is maximum. It is difficult to find any systematic
principle or procedure that can optimize the efficiency. 

Transport properties in overdamped inhomogeneous systems   
have been dealt with great detail 
previously~\cite{But,Parek,Spc_mob}. Occurrence of multiple current
reversals~\cite{dan2}, current reversal 
under adiabatic or deterministic conditions, unidirectional
motion in the absence of potential ~\cite{Spc_mob}, stochastic resonance in the
absence of periodic forcing have also been observed~\cite{dan1}. Most of these phenomena
arise solely due to the presence of frictional inhomogeneity. It is to be
noted that systems with space dependent friction are not uncommon. 
Brownian motion in confined geometries show space dependent
friction. Particles diffusing close to surface have space dependent
friction coefficient~\cite{dan2}. It is also believed that molecular
motors move 
close along the periodic structure of microtubules and will therefore
experience a position dependent mobility~\cite{Spc_mob}. Frictional
inhomogeneities are common in super lattice structures and
semiconductor systems~\cite{But}.    

                 We consider an overdamped Brownian particle moving in 
an inhomogeneous 1D ratchet like potential, rocked by a finite frequency 
driving force. We consider an asymmetric potential of the form  $V(x)
= -1/(2\pi) (\sin(2\pi x) + \mu/4 \sin(4\pi x)) + Lx$, where $L$ is
the external load against which the Brownian particle moves on
average. $\mu$ is the asymmetry parameter and is in between the range
$0$ and $1$. The direction of load is chosen against the mean drift of
the Brownian particle so that the work done by the particle
is positive. The system is rocked by a zero mean external
force of the form $F(t) = A \sin (\omega t)$. The correct Langevin equation
for such a motion has been derived using microscopic treatment of
system bath coupling ~\cite{Parek,Sancho}.  
\begin{equation}
  \label{Langvn}
  \dot{x} =  -\frac{(V'(x) - F(t))}{\eta (x)} - k_{B}T \frac{\eta '(x)}{(\eta
   (x))^{2}} + \sqrt{\frac{k_{B}T}{\eta (x)}}\xi (t) ,
\end{equation}
The quantity $x$ represents the spatial position
of the system. It should be
noted that the above equation involves a multiplicative noise with an
additional temperature dependent drift term which turns out to be
essential for the system to approach correct thermal equilibrium
state in absence of external drive $F(t)$ and load $L$
~\cite{Parek,Sancho}. The Gaussian  
white noise $\xi(t)$ is delta correlated with mean zero, 
i.e., $< \xi(t) \xi(t^{\prime})> = 2D\delta(t-t^{\prime})$. 
The friction coefficient $\eta (x) = \eta_{0}(1-\lambda \sin(2\pi x + \phi))$, 
$|\lambda| < 1$ and $\phi$ determines the relative
phase shift between  friction coefficient and potential.  The Fokker
Planck equation corresponding to eqn. (~\ref{Langvn}) is given by ~\cite{Ris}
\begin{equation}
 \frac{\partial P(x,t)}{\partial t} = -\frac{\partial J(x,t)}{\partial 
   x} = \frac{\partial}{\partial x}
 \frac{1}{\eta (x)} [k_{B}T \frac{\partial}{\partial x} + (V'(x) -
 F(t))] P(x,t) ,
 \label{FPE}
\end{equation}
where $J(x,t)$ and $P(x,t)$ are the current density and probability
density respectively. The mean current
\begin{equation} 
  \label{current}
  J = \lim_{t \rightarrow \infty} \frac{1}{\tau} \int_{t}^{t+\tau} dt
 \int_{0}^{1} J(x,t) dx ,
\end{equation}
which is obtained numerically by solving eqn. (\ref{FPE}) by the 
method of finite difference. 
 The work done against the load, 
given by $W = L J$. 
The average input energy $E$ is given by $E = \lim_{t \rightarrow \infty}
\frac{1}{\tau}\int_{t}^{t+\tau}dt \int_{0}^{1} 
F(t) J(x,t) dx$.
The efficiency of the
system to transform the external force to useful work (storing 
potential energy) is ~\cite{taka1}
\begin{equation}
  \label{effi}
   \eta = \frac{W}{E} = \frac{LJ}{\lim_{t \rightarrow \infty}
\frac{1}{\tau}\int_{t}^{t+\tau}dt \int_{0}^{1} 
F(t) J(x,t) dx} ,
\end{equation}
where $J$ is calculated from eqn. (\ref{current}).

        We now discuss the effect of finite frequency
drive, spatial asymmetry and inhomogeneous friction coefficient on the
efficiency of energy transduction. It is observed that spatial asymmetry or
space dependent friction coefficient alone cannot enhance the
nonadiabatic efficiency as compared to the adiabatic one 
in a rocked thermal ratchet. However, the interplay of
both can enhance the efficiency in the nonadiabatic regime.  

\begin{figure}
   \centerline{\epsfysize=18cm \epsfbox{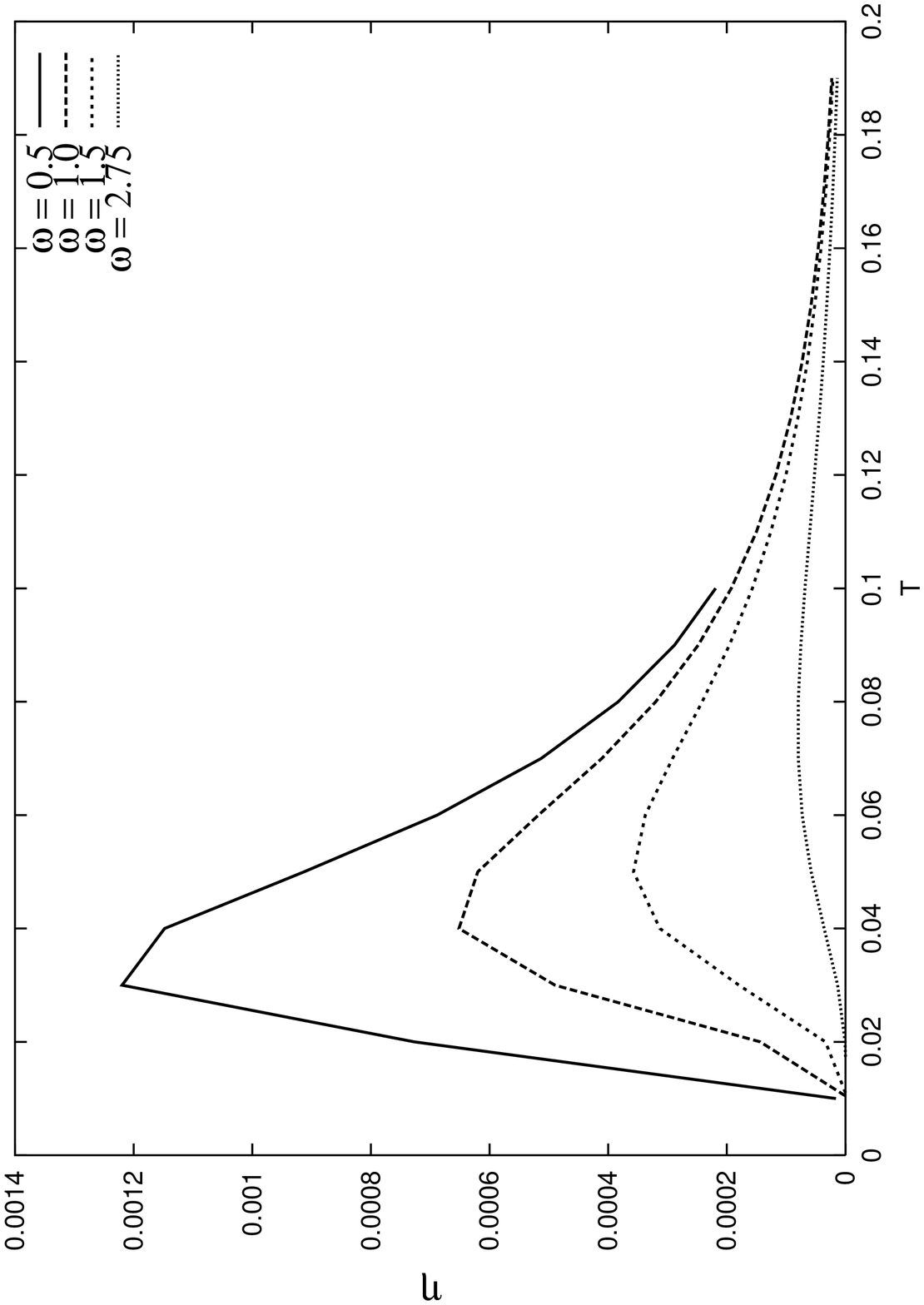}}
   \caption{\large Efficiency $\eta$ vs temperature $T$ for $A=0.5, \mu=1.0,
     \lambda=0, L=0.001$ and various values of 
     $\omega$. The curve for $\omega = 0.5$ follows the same trend
     as other curves beyond $T = 0.1$}
   \label{fig1a}
 \end{figure}

    First we discuss the nature of efficiency in an asymmetric ratchet 
in the absence of spatial frictional inhomogeneity ($\lambda = 0, \mu = 1$).
This ratchet in contact with thermal bath produces
 directed motion when rocked by a finite force. The direction of
current being dependent both on the direction of asymmetry as well as the
frequency of the driving force ~\cite{bartu}. 
When rocked adiabatically, the current 
shows maxima at some nonzero value of temperature. Even though the
current shows a maxima the efficiency
monotonically decreases with temperature ~\cite{taka1,Dan3}. 
The situation changes
in the nonadiabatic regime as shown in the
fig.~\ref{fig1a}. Throughout this work temperature and frequency have been 
scaled appropriately to make them dimensionless~\cite{Ris}. 
In fig.~\ref{fig1a} we have plotted $\eta$ vs $T$ for
various values of $\omega$. It can be seen that unlike the adiabatic
case, $\eta$ shows a maxima at a nonzero value of temperature. The
value at which $\eta$ peaks, decreases with decreasing frequency, as it
should be. We have observed that even though the efficiency peaks 
at nonzero value of
temperature in the nonadiabatic regime, efficiency in the adiabatic
regime (at $T=0$)is
much larger than the peak nonadiabatic efficiency.

\begin{figure}
   \centerline{\epsfbox{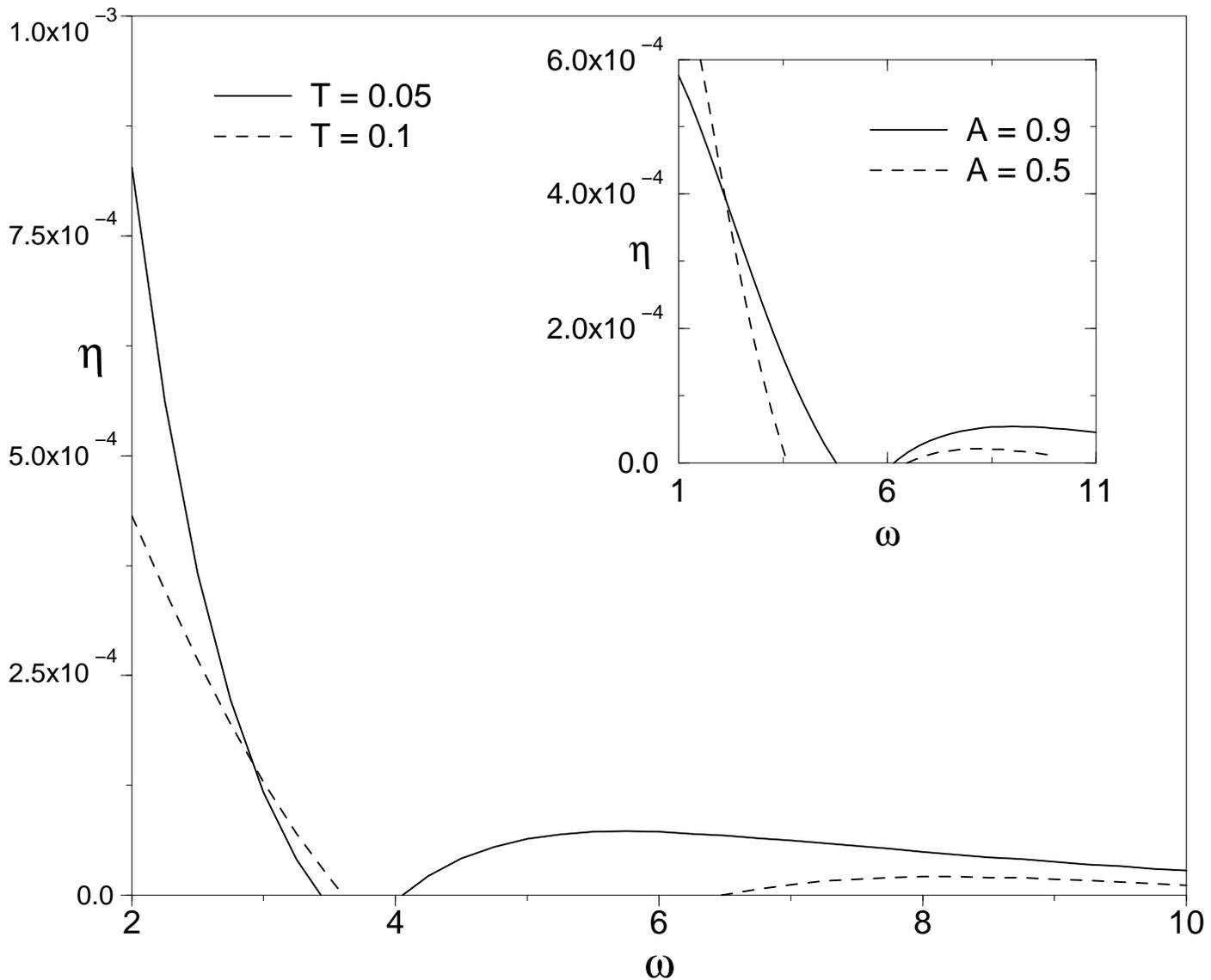}}
   \caption{\large Efficiency vs $\omega$ for two values $T$ at $A =
     0.5, \mu=1.0, \lambda=0.0$ and $|L| = 0.005$. The inset shows
     variation of $\eta$ with $\omega$ for two values of $A$ at
     $T=0.1$, all other parameter values remaining same.}
   \label{fig1b}
 \end{figure}

            In fig.~\ref{fig1b} we have plotted efficiency as 
function of  
rocking frequency for various values of $T$ (and $A$, in the
inset). Current reversal as  function of frequency is a common 
phenomena in a driven asymmetric ratchet.  
Since beyond a critical frequency, the current reverses its
direction \cite{bartu}, the load has been applied in the opposite direction 
in that
regime so that work is done against the load.   
As shown in the fig.~\ref{fig1b}, for
low frequencies, efficiency shows a monotonic decrease with
frequency. The rate of decrease of $\eta$ with $\omega$ being
critically dependent on temperature $T$ and amplitude $A$. 
In the current reversed regime, the efficiency shows
a maxima with $\omega$, though its value is much less than the
adiabatic efficiency. We have verified this fact by exhaustive
numerical work with our given potential.

\begin{figure}
   \centerline{\epsfysize=18cm \epsfbox{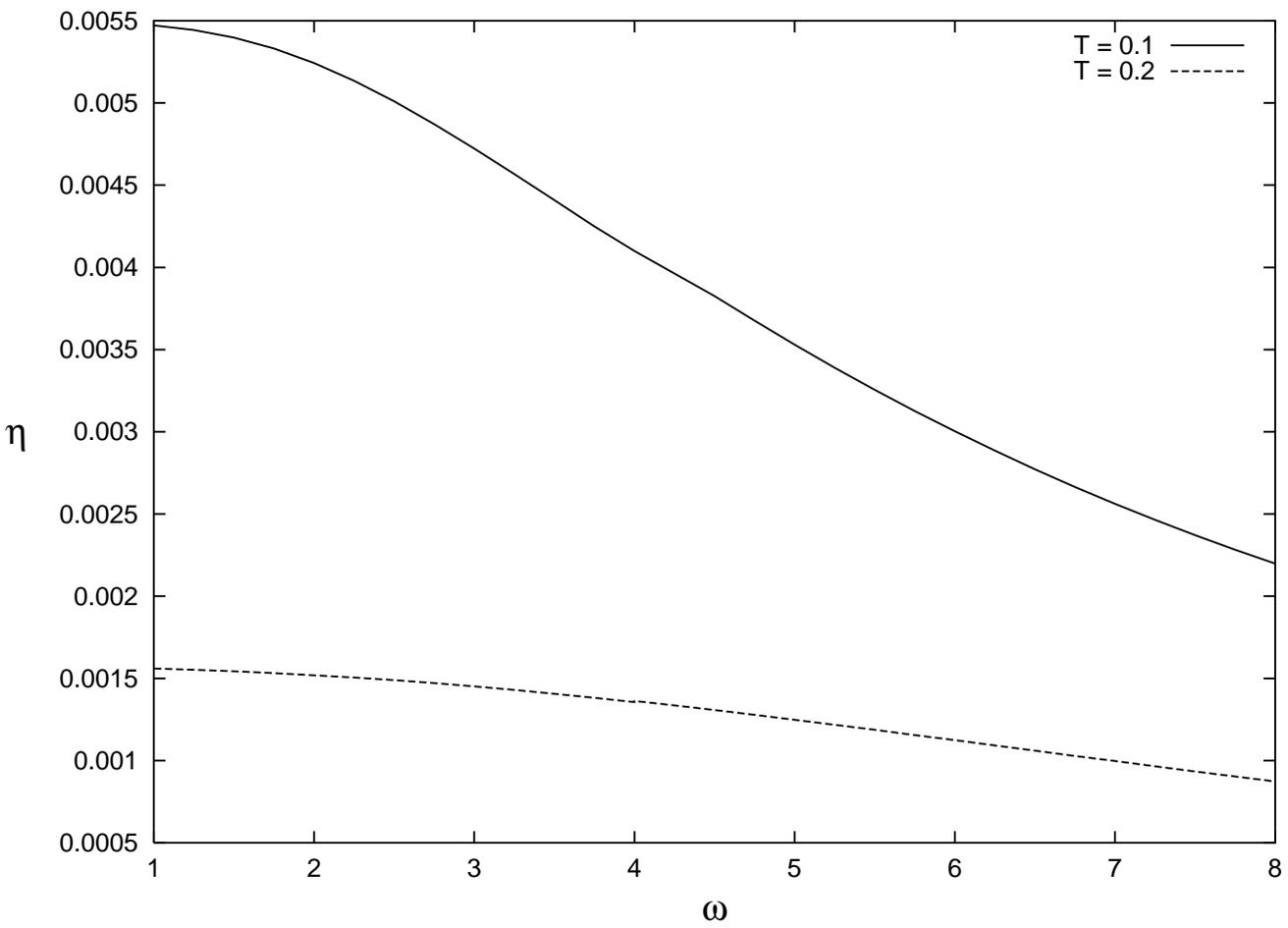}}
   \caption{\large Efficiency vs $\omega$ for $\mu=0, \lambda=0.9,
     \phi=0.6\pi, L=-0.012$ and for two values of $T$}
   \label{fig2}
 \end{figure}

              We now consider a system in which friction is space dependent 
with a \textit{symmetric}  potential profile. Unidirectional 
current results whenever $\phi \neq 0, \pi$ or $2\pi$ as discussed in 
ref. (~\cite{Dan3}). In these models inversion symmetry is broken 
dynamically by space dependent friction.
This system does not exhibit current reversal with any of the
variables like $T, A$ or $\omega$ (in the absence of $L$) and hence we 
keep the load fixed in one direction for comparison of  efficiency. In
fig.~\ref{fig2} we have plotted $\eta$ \textit{vs} $\omega$ for $A=0.5 ,
\phi=0.6 \pi $ and $L = -0.012 $. For all values of $T,\lambda$ and
$\phi$, the efficiency 
monotonically decreases with $\omega$, i.e., for a given $T$ the
adiabatic efficiency is always maximum. In the two cases considered
above ($\lambda=0$ with asymmetric potential and $\lambda \neq 0$ with 
symmetric potential ) adiabatic current is always more than the
absolute value of the peak current in the nonadiabatic regime. The efficiency
in our present case is mainly determined by the nature of currents and 
hence the result follows.

\begin{figure}
   \centerline{\epsfbox{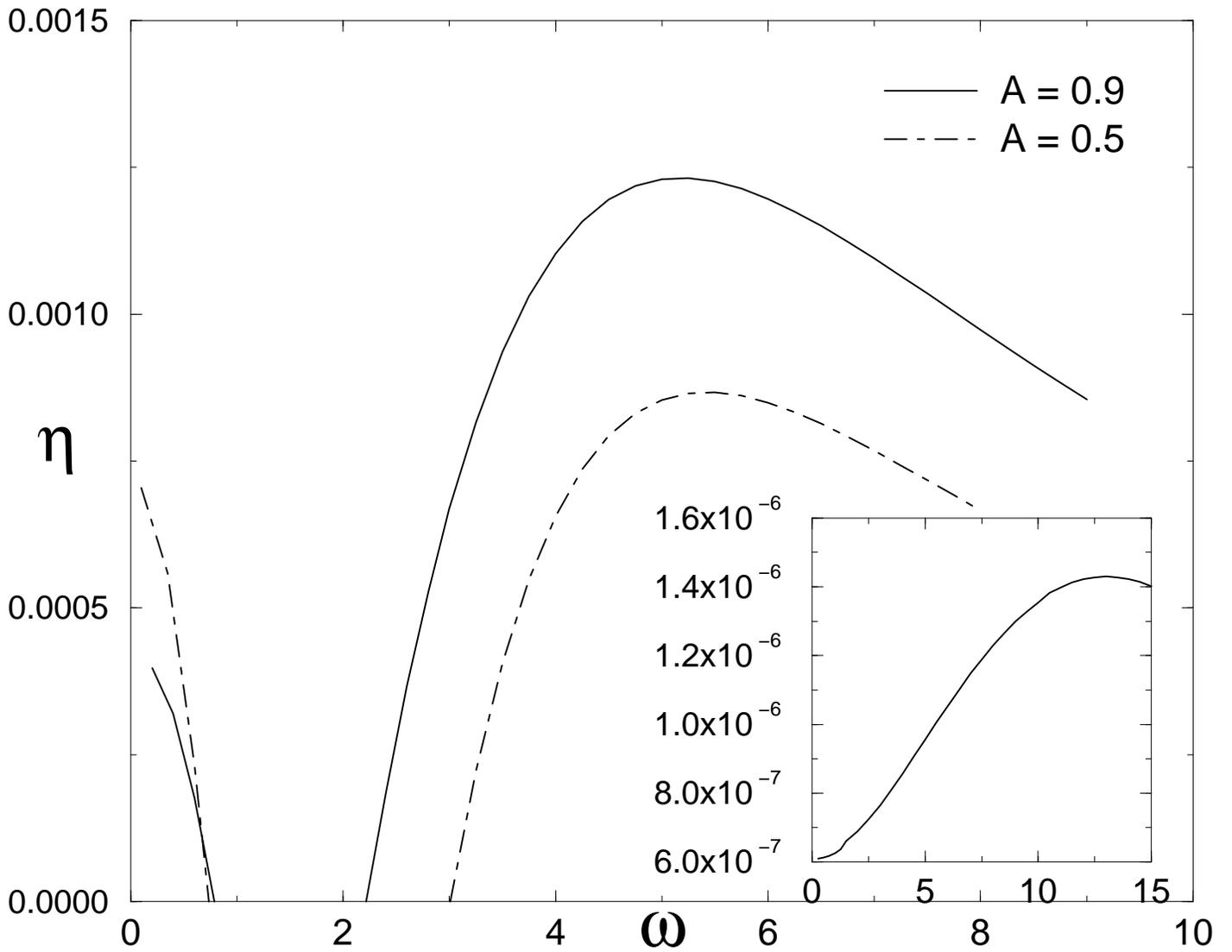}}
   \caption{\large Efficiency vs $\omega$ for two values of $A$. For other
     parameter values see text. The inset
     shows the variation of $\eta$ vs $\omega$ for $A=1.5,
     T=0.4$. }
   \label{fig3a}
 \end{figure}

             We now concentrate on frictional ratchets with asymmetric
potential profile ($\lambda \neq 0, \mu \neq 0$).  
  The efficiency characteristics of these ratchets
have many novel and counterintuitive features. In
fig.~\ref{fig3a} we plot  
efficiency as function of
$\omega$ for two values of forcing amplitude $A$ with $T = 0.08,
\lambda = 0.9, \phi = 0.2 \pi$ and $L = 0.015$. 
It can be clearly seen that the nonadiabatic efficiency is
higher than the adiabatic efficiency, which is contrary to common
belief that a rocked Brownian ratchet is inefficient in the nonadiabatic 
domain. This enhanced efficiency basically results from the \textit{increase
in the current with increase of
frequency in current reversed regime}. The increase of current in this
regime can be
ascribed due to mutual 
interplay of spatial asymmetry and space dependent friction
coefficient ~\cite{Dan3}. The phase difference $\phi$ is chosen in
such a manner 
so that the steeper side has lower friction coefficient than the
slanted side. The inset shows a
different qualitative behavior of $\eta$ with increase of $\omega$.
Here $A = 1.5, T = 0.4, \lambda=0.1, \phi = 0.2\pi$ and $L = -0.001$. In this
parameter regime there is no current reversal. Here as we increase
frequency from adiabatic regime, $\eta$ increases till it exhibits a
maxima at very high frequency and decreases on further increasing the
frequency.At high $T$ and in the adiabatic regime, particles get sufficient
kicks and enough time to cross the barrier on both the side, but the
frictional drag on the steeper side is less. Hence the current flow is 
in the negative direction. On increasing the frequency, the Brownian
particles get less time to cross the right barrier as it has to travel 
larger distance to reach the basin of attraction of the next well than 
from the left side. Hence the net current increases. From the above
argument it can be easily seen that the efficiency
increases with increasing asymmetry (increasing $\mu$) and vice-versa which we have checked
in our work. For too high
frequencies the particles do not get sufficient time to cross either
of the barriers and the current decreases which reflects in the
decrease of efficiency as shown in the fig.~\ref{fig3a}. Hence efficiency
optimization at high frequency is not only  
a phenomenon in current reversed regime but other wise also. 
 On decreasing temperature the asymmetric ratchet
effect becomes more pronounced and efficiency
increases along with the shift of the peak efficiency to lower
frequency regime.  
  
\begin{figure}
   \centerline{\epsfbox{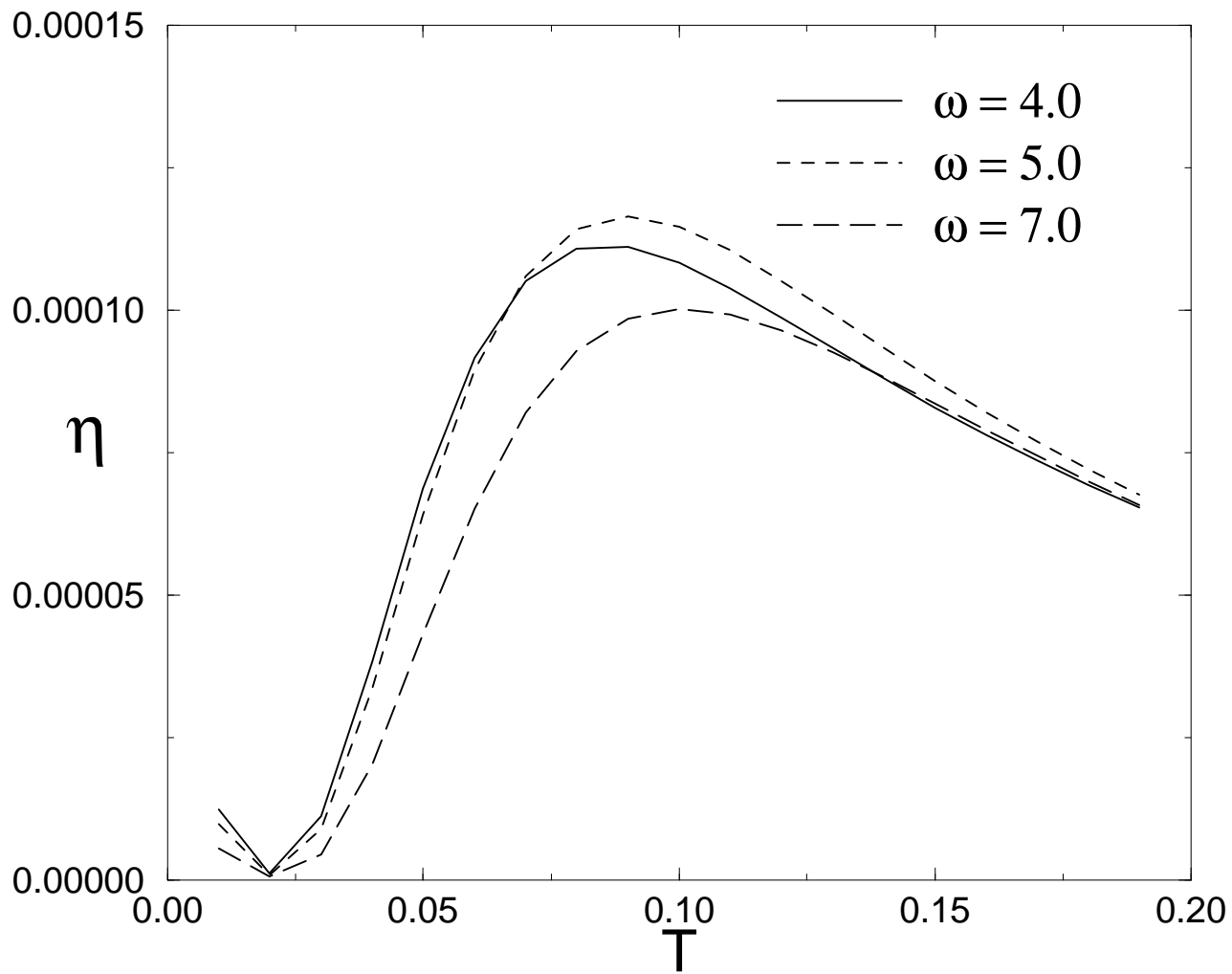}}
   \caption{\large Efficiency vs temperature for different values of
     $\omega$. Here $\mu=1, \lambda=0.9, A=0.5, \phi=0.2\pi$ and $L=-0.001$ }
   \label{effi-T}
 \end{figure}

Like other previous cases depending on the parameter values the
efficiency ( as a function of $T$) 
may or may not be maximized  at finite temperature ~\cite{taka1,Dan3}. 
This optimum value (if maxima exists) increases initially with
increasing frequency 
and then for too high frequencies it decreases as discussed
earlier. The temperature at which $\eta$ peaks increases
with increasing frequency as shown in the fig.~\ref{effi-T}. This
shows that unlike conventional wisdom where we associate high driving
frequency and temperature with inefficient energy conversion, here
both \textit{high frequency} (nonadiabatic regime) and
\textit{temperature enhances efficiency}.


                 In conclusion we have studied the efficiency of
energy transduction in a forced frictional ratchet as function of
rocking frequency. Both
nonmonotonic and monotonic behavior have been observed. In particular 
the magnitude of efficiency in finite frequency regime may be more
than the efficiency in the adiabatic regime. This implies that in
these rocked ratchet systems quasi-static operation may not be
efficient for conversion of input energy into mechanical work. 
Observation of peak in the efficiency as function of system
parameters can be qualitatively 
attributed to the peak in the current and not to the behavior of
input energy, though the occurrence of peak in current may not 
guarantee a peak in efficiency as observed earlier
~\cite{taka1,Dan3}. Here we have taken a simple ratchet type potential 
with space dependent friction coefficient 
to illustrate the above results.  We 
do not rule out the fact that similar result can also be obtained in
homogeneous systems for  different ratchet potentials provided they exhibit
larger absolute peak current in the nonadiabatic regime.  It is interesting 
to explore this possibility. It should be noted that in flashing ratchets
unlike rocking ratchets, efficiency can show peaking behaviour as a 
function of frequency. This is because both in zero frequency and in
high frequency limit flashing ratchet does not exhibit current ~\cite{ajdari}.  
Detailed
study of input energy, work done and dependence of efficiency on other 
system parameters will be reported in future.

\end{document}